\documentclass[12pt]{article}
\usepackage{amssymb}

\usepackage[dvips]{epsfig}

\begin{document}

\author{Augusto E. R. Chumbes, Marcelo B. Hott \thanks{
e-mail: hott@feg.unesp.br} \\
\\
Universidade Estadual Paulista\\
Departamento de F\'{\i}sica e Qu\'{\i}mica\\
12516-410. Guaratinguet\'{a}, SP - Brasil}
\title{{\Large Non-polynomial potentials with deformable topological
structures}}
\maketitle

\begin{abstract}
We construct models of self-interacting scalar fields whose BPS solutions
exhibit kink profiles which can be continuously deformed into two-kinks by
varying one of the parameters of the self-interacting potential. The
effective models are obtained from other models with two interacting scalar
fields. The effective models are then applied in a brane-world scenario
where we analyze the consequences of the thicker branes in the warped
geometry and in the localization of gravity.
\end{abstract}

\section{Introduction}

Nonlinear systems, particularly those that exhibit topological solutions are
very important for modeling many physical systems from condensed matter to
high-energy physics. One example that has\ been receiving considerable
attention is that of topological structures in multidimensional warped
space-time when one considers scalar fields coupled to gravity. In this
context, one of the key points is the possibility of localizing gravitons in
thin branes and reproducing effectively the four-dimensional gravity \cite%
{randallsundrum}. In (4,1) dimensions it is shown that thick branes can also
localize the gravity \cite{gremm}. In fact, the subject of thick branes in
the context of brane-worlds has received a considerable amount of attention
\cite{kyrgiz}. Some years ago it was observed that some kinds of models with
two interacting scalar fields in a warped geometry can be used to describe
the splitting of thick branes due to a first-order phase transition \cite%
{campos}. A few years ago, it was shown that Bloch branes are solutions of a
model with two interacting scalar fields which can be naturally incorporated
in a supersymmetric theory \cite{bazeia1}. Later, that same model was shown
to exhibit branes with richer structures \cite{nosso} than those found in
the reference \cite{bazeia1}, including critical and degenerate branes.
Moreover, contrary to the scenarios presented in\ \cite{campos} and \cite%
{bazeia1} where the splitting of the branes is controlled by a coupling
constant presented in the interaction potential, it was shown \cite{nosso}
that the thickness of the branes is controlled by a parameter, called
\textit{degeneracy parameter, }which is not present in the Lagrangian
density of the model. Instead, it is one of the constants of integration of
the orbit differential equation which relates both fields. Recently, a model
with only one scalar field which incorporates thicker branes was proposed by
Dutra \cite{dutra09}. It is characterized by a non-polynomial interaction
potential with coupling constants that control the thickness of the brane.
This last property, besides the fact of exhibiting a non-polynomial
interacting potential, is shared with the \textit{p-model} introduced in the
references \cite{bazeia2} and \cite{bazeia3}. In the later one, the changing
of the parameter \textit{p }in discrete jumps implies into a changing of
model and thicker branes appears only for some values of \textit{p},
whereas, in the model introduced in \cite{dutra09}, the parameter that
controls the deformation of the brane is a coupling constant of the model
and as such, its variation do not modify the structure of the model.

The purpose of this work is to construct nonlinear models, in classical
field theory, with only one scalar field,\ from models with two interacting
scalar fields and that exhibit solutions with \textit{two-kinks} profiles.
Two-kinks solutions yield thicker branes in brane-worlds scenarios or Bloch
branes whose internal structures are somehow incorporated in the parameter
that controls the thickness of the brane. The non-polynomials effective
models we construct belong to the same class of models than that proposed by
Dutra \cite{dutra09}, where the kink solutions can be continuously
deformable into two-kinks by varying one of the parameters of the effective
potential. Moreover, the same parameter that controls the thickness of the
brane can be thought as depending on the temperature in such a way that a
first-order phase transition characterized by the brane splitting can
happen. The model with two interacting scalar fields we consider here is the
same whose consequences in brane-worlds scenarios were already studied in
the references \cite{bazeia1} and \cite{nosso}. In order to construct the
effective models, we apply a general orbit equation relating both fields
\cite{dutra05} to eliminate one of the fields in favor of the other. The
resulting effective model is applied to the study of the localization of
gravity by thick branes in (4,1)-dimensional warped space-time. In the next
section we introduce the kind of models with two interacting fields we are
interested in and, after that, we show how the orbit equation can be used to
construct models with only one scalar field. One particular example is
studied in detail. In the third section we apply one of the effective models
in brane-world scenario and analyze the consequences of thick branes over
the warp factor and on the localization of gravity. In the conclusions
section we remark on possible applications of the effective models,
particularly in the phenomenon of brane splitting.

\section{Models with only one scalar field constructed from models with two
interacting scalar fields}

The models with two interacting fields we consider here are described by the
Lagrangian density
\begin{equation}
{\mathcal{L}}=\frac{1}{2}\partial ^{\mu }\phi \partial _{\mu }\phi +\frac{1}{%
2}\partial ^{\mu }\chi \partial _{\mu }\chi -V(\phi ,\chi ),  \label{eq1}
\end{equation}%
with the potential written as
\begin{equation}
V(\phi ,\chi )=\frac{1}{2}(W_{\phi }^{2}+W_{\chi }^{2}).  \label{eq2}
\end{equation}%
where $W_{\phi }$ and $W_{\chi }$ are the derivatives of some function $%
W(\phi ,\chi )$, called superpotential, with respect to the fields $\phi $
and $\chi $, respectively.

It is shown that for potentials written in terms of a superpotential, as in
equation (\ref{eq2}), the static solutions of the first-order equations
\begin{equation}
\frac{d\phi }{dx}=W_{\phi },~~\frac{d\chi }{dx}=W_{\chi },  \label{eq3}
\end{equation}%
are those that minimize the energy of the system, the BPS energy \cite{BPS},
and are also solutions of the static equations of motions \cite{bazeia0}.
Based on the first-order differential equations one can realize that $dx$ is
a kind of invariant and the following, in general nonlinear equation
\begin{equation}
\frac{d\chi }{d\phi }=\frac{W_{\chi }}{W_{\phi }}  \label{eq4}
\end{equation}%
furnishes a relation between the classical static solutions, called the
orbit equation, which can be solved analytically depending on the model
under consideration \cite{dutra05}. For the cases in which the general orbit
equation can be found, we can write the scalar field $\chi $ in terms of the
scalar field $\phi $, that is $\chi =f(\phi )$, and it can be used to
eliminate $\chi $ in terms of $\phi $ in the Lagrangian density (\ref{eq1}),
that is
\begin{equation}
{\mathcal{L}}=\frac{1}{2}(1+f_{\phi }^{2})\partial ^{\mu }\phi \partial
_{\mu }\phi -V(\phi ),  \label{eq5}
\end{equation}%
where $f_{\phi }=\frac{df}{d\phi }=\frac{d\chi }{d\phi }=\frac{W_{\chi }}{%
W_{\phi }}$ with both, $W_{\phi }$ and $W_{\chi }$, written in terms only of
the field $\phi $. $V(\phi )$, is the potential $V(\phi ,\chi )$ with $\chi $
written in terms of $\phi $. Due to the peculiar structure of the potential
given in (\ref{eq2}) and also to the differential equation (\ref{eq4}), we
have
\begin{equation}
V(\phi )=\frac{1}{2}(1+f_{\phi }^{2})W_{\phi }^{2}  \label{eq5a}
\end{equation}%
\

The equation of motion for this single scalar field model is given by
\begin{equation}
(1+f_{\phi }^{2})\partial _{\mu }\partial ^{\mu }\phi +f_{\phi }f_{\phi \phi
}\partial _{\mu }\phi \partial ^{\mu }\phi +\frac{dV}{d\phi }=0,  \label{eq6}
\end{equation}%
and the energy associated with the static classical solutions is expressed
as
\begin{equation}
E=\int dx({\frac{1}{2}(1+f_{\phi }^{2})\phi ^{\prime }{}^{2}+V(\phi )}),
\label{eq7}
\end{equation}%
where ${\phi ^{\prime }}$ stands for $d\phi /dx.$In order to find the
minimum energy we note that the energy can be rewritten as
\begin{equation}
E=\frac{1}{2}\int dx\left[ ((1+f_{\phi }^{2})^{1/2}\phi ^{\prime }\pm \sqrt{%
2V})^{2}\mp 2(1+f_{\phi }^{2})^{1/2}\phi ^{\prime }\sqrt{2V}\right] ,
\label{eq8}
\end{equation}%
and, consequently, the classical solutions with minimum energy satisfies the
first-order differential equations
\begin{equation}
\phi ^{\prime }=\mp (\frac{2V}{1+f_{\phi }^{2}})^{1/2}=\mp W_{\phi },
\label{eq9}
\end{equation}%
where $W_{\phi }=dW(\phi ,\chi )/d\phi \,\ $with $\chi $ replaced by $f(\phi
)$. It is easy to show that the solutions of the first-order differential
equations are also solutions of the static equation of motion
\begin{equation}
(1+f_{\phi }^{2})\phi ^{\prime \prime }+f_{\phi }f_{\phi \phi }\phi ^{\prime
^{2}}=\frac{dV}{d\phi }.  \label{eq10}
\end{equation}%
and the BPS energy is given by
\begin{eqnarray}
E_{BPS} &=&\int dx((1+f_{\phi }^{2})\phi ^{\prime }W_{\phi }(\phi ,f(\phi
)))=  \nonumber \\
&=&|W(\bar{\phi},f(\bar{\phi}))(\infty )-W(\bar{\phi},f(\bar{\phi}))(-\infty
)|,  \label{eq11}
\end{eqnarray}%
where the classical solutions $\bar{\phi}(x)$ are to be taken at $\pm \infty
$.

This procedure resembles the one carried out in the reference \cite{dutra04}
to prove the equivalence between sine-Gordon, Liouville and other models. In
that reference the first-order differential equations obeyed by the BPS
solutions of \ the models are used to construct a mapping between the fields
of the two models whose equivalence is to be demonstrated. In fact, one
deforms one known model by using the mapping function and obtains another
known model. It is shown that both models possess the same BPS energy when
the deformation is performed in the Lagrangian density, in contrast to what
happens when the deformation is carried out in the differential equations
\cite{bazeia4}. One can see that the BPS energy found in equation (\ref{eq11}%
) of the model given by the Lagrangian density (\ref{eq5}) is the same of
the model described by (\ref{eq1}). Since we are not interested in proving
the equivalence between those models we would rather prefer to work with an
effective Lagrangian density whose structure is of the type kinetic$-$%
potential which leads to static Euler-Lagrangian equation of the type $\phi
^{\prime \prime }=dU_{eff}/d\phi $, whose classical solutions are also
solutions of the equation (\ref{eq10}). This procedure is more like the one
of \cite{bazeia4} and \ it is more convenient when one works with scalar
fields in interaction with gravitation as we consider in the next section.

\subsection{Particular cases:}

Here we consider the same model with two interacting scalar fields applied
in \textit{brane worlds} scenario in the references \cite{bazeia1} and \cite%
{nosso} whose superpotential is
\begin{equation}
W(\phi ,\chi )=\phi \lbrack \lambda (\frac{\phi ^{2}}{3}-a^{2})+\mu \chi
^{2}].  \label{eq12}
\end{equation}

In a recent paper, Dutra \cite{dutra05} was able to show that for this case
the orbit equations relating both fields can be obtained explicitly. They
can be written as
\begin{equation}
\rho (\chi )=\phi ^{2}-a^{2}=c_{0}\chi ^{\lambda /\mu }-\frac{\mu }{\lambda
-2\mu }\chi ^{2},\hspace{0.5in}\mathrm{for~}\lambda \neq 2\mu  \label{eq13}
\end{equation}%
and
\begin{equation}
\rho (\chi )=\phi ^{2}-a^{2}=\chi ^{2}[\ln (\chi )+c_{1}],~\hspace{0.5in}%
\mathrm{for~}\lambda =2\mu ,  \label{eq14}
\end{equation}%
where $c_{0}$ and $c_{1}$ are constants of integration. In general, only the
first of the above orbit equations is used to find well behaved classical
solutions. One can check, for example, that the second of the above orbit
equation fails to reproduce the minima of the model, namely $\phi =\pm a$
and $\chi =0$ . From now one we consider $a>0$.

The first orbit equation can be used to construct models with only one
scalar field which exhibit the main features of this model with two
interacting scalar fields. An one example we consider the situation in which
$\mu =\lambda $. In this case the orbit equation is given by%
\begin{equation}
\chi ^{2}+c_{0}\chi -(\phi ^{2}-a^{2})=0,  \label{eq15}
\end{equation}%
and the field $\chi $ can be expressed in terms of $\phi \,\ $as
\begin{equation}
\chi =f(\phi )=-\frac{c_{0}}{2}\pm \frac{1}{2}\sqrt{c_{0}^{2}+4(\phi
^{2}-a^{2})}.  \label{eq15a}
\end{equation}%
By substituting the above expression with the upper sign in the expression
for $W_{\phi }(\phi ,\chi =f(\phi ))$ we obtain the following superpotential
\begin{equation}
{W}_{\phi }(\phi )=2\mu (\phi ^{2}+b^{2}-a^{2}+b\sqrt{\phi ^{2}+b^{2}-a^{2}}%
).  \label{eq17}
\end{equation}

From now on we can work with a model described by the Lagrangian density
\begin{equation}
{\mathcal{L}}=\frac{1}{2}\partial ^{\mu }\phi \partial _{\mu }\phi
-U_{eff}(\phi ),  \label{eq17a}
\end{equation}%
with the effective potential $U_{eff}(\phi )=W_{\phi }^{2}(\phi )/2$. In
this model the first-order differential equations satisfied by $\phi (x)$ is
$\phi ^{\prime }=\mp W_{\phi }$ and the solutions for the field $\phi $ are
the same obtained in the model with two interacting scalar fields \cite%
{nosso, dutra05} described by the Lagrangian density (\ref{eq1}) with $%
W(\phi ,\chi )$ given in (\ref{eq12}).

The choice of the lower sign in the expression (\ref{eq15a}) would result in
an effective potential $U_{eff}(\phi )=W_{\phi }^{2}/2$ with only one
minimum and we are interested in effective potentials with at least two
minima. The constant $b=c_{0}/2$ must satisfy the inequality $b<-a$, such
that we have non-singular solutions and the effective potential presents two
global minima and one local minimum for a certain range of the parameter $b$%
. On the other hand, for $b=-a$, we have an effective potential with three
global minima. We show in the Figure 1 the behavior of the effective
potential, in units of \ $\mu ^{2}$, as a function of $\phi $, for two
different values of $b$ and $a=1$. One can see that for $b<-a$, the
effective potential tends to exhibit three minima as $b$ gets closer to $-a$%
. In fact, for $b=-a$ the effective potential becomes $U_{eff}(\phi )=2\mu
^{2}(\phi ^{2}-a|\phi |)^{2}\,$that has also a minimum at $\phi =0$. The
classical solutions of the first-order differential equation (\ref{eq9}),
for $b<-a,$ are given by
\begin{equation}
\phi =\pm a\frac{\sinh (2\mu ax)}{\cosh (2\mu ax)-b/f}~,  \label{eq18}
\end{equation}%
where $f=\sqrt{b^{2}-a^{2}}$ and the upper (lower) sign stands for the kink
(anti-kink) solution. In the Figure 2 it is shown a profile of the
topological classical solution for a sufficiently large value of \ $|b|$,
and a double kink profile, usually called two-kinks, for $b$ close to the
critical value $-a$. This\ kind of configuration also arises as solution of
the models introduced in the references \cite{dutra09} and \cite{bazeia2}.

Another choice of the parameter $\lambda $, namely $\lambda =4\mu $, leads
to a similar non-polynomial effective potential. In this case, the orbit
equation can be written as
\begin{equation}
\chi ^{2}=\frac{1}{4c_{0}}\left[ 1\pm \sqrt{1+16c_{0}(\phi ^{2}-a^{2})}%
\right] ,  \label{eq19}
\end{equation}%
and, by taking the upper sign in the above equation, we have the following
effective superpotential
\begin{equation}
W_{\phi }=4\mu \left( \phi ^{2}-d^{2}+dc\sqrt{\phi ^{2}-d^{2}}\right) ,
\label{eq20}
\end{equation}%
where $d=a/\sqrt{c^{2}-1}$ and $c=1/\sqrt{1-16c_{0}a^{2}}$. In order to have
a well defined model, the constant $c_{0}$ must satisfy the inequality $%
c_{0}<1/16a^{2}$ and the solutions are given by
\begin{equation}
\phi =\pm a\frac{\sinh (4\mu ax)}{\cosh (4\mu ax)+c}.  \label{eq21}
\end{equation}

It worth mentioning that the model we have constructed\ in this case ($%
\lambda =4\mu $) is very similar to the one proposed in the reference \cite%
{dutra09} if we set $\mu =1/4$ and make some identification between the
parameter $d$ with the parameter $b_{0}~$from that paper. The behavior of
the effective potential as a function of $\phi $ is almost identical to the
one presented in the previous case. Moreover, if $c_{0}=1/16a^{2}$ one
obtains the effective potential, $U_{eff}(\phi )=8\mu ^{2}(\phi ^{2}-a|\phi
|)^{2}$,$\,$ with three minima.

Effective polynomial potentials can also be obtained from the model
described by the Lagrangian density (\ref{eq1}) with $W(\phi ,\chi )$ given
by (\ref{eq12}). This is done by using the orbit equation (\ref{eq13}) to
express the field $\phi $ as a function of \ the field $\chi $. By
conveniently rewriting the orbit equation as
\begin{equation}
\phi =g(\chi )=\pm \sqrt{c_{0}\chi ^{\lambda /\mu }-\frac{\mu }{\lambda
-2\mu }\chi ^{2}+a^{2}},  \label{eq22}
\end{equation}%
we have the effective potential given by
\begin{equation}
U_{eff}(\chi )=\frac{1}{2}W_{\chi }^{2}=2\mu ^{2}\chi ^{2}\left( c_{0}\chi
^{\lambda /\mu }-\frac{\mu }{\lambda -2\mu }\chi ^{2}+a^{2}\right) .
\label{eq23}
\end{equation}%
\qquad

This potential has at least one minimum at $\chi =0$, and we have to set $%
\lambda /\mu =n$, where $n$ is a positive integer ($n\neq 2$), in order to
have well defined potentials with, at least, two minima. This last condition
leads to some constraints over the values of $c_{0}$. We note that for $%
\lambda /\mu =1$, \ a positive definite potential with two minima, is
obtained if $c_{0}=\pm 2a$ and, for $\lambda /\mu =4,$ we find a positive
definite potential with three minima at $\chi =0,\pm 2a$ if $c_{0}=1/16a^{2}$%
. Those specific values for $c_{0}$ are the critical ones that lead to
polynomials models with respect to the field $\phi $ which possess three
minima. We look for the critical values of $c_{0}$, corresponding to
different values of $n$, in order to have positive definite effective
potentials, $U_{eff}(\chi )$, with more than one minimum. First, we note
that effective potentials of the type $U_{eff}(\phi )=(\phi ^{2}-a|\phi
|)^{2}$ has classical solutions that connect the minimum $\phi =0$ to the
minima $\phi =+a$ or $\phi =-a$ and vice-versa. Moreover, we recall that
those polynomial potentials, are constructed by substituting the orbit
equation $\chi =f(\phi )$ into the superpotential $W_{\phi }=\lambda (\phi
^{2}-a^{2})+\mu \chi ^{2}$ and this one, by its turn, is substituted in $%
U_{eff}(\phi )=W_{\phi }^{2}/2$, which is positive definite everywhere. If $%
\phi =0$ is a minimum of this effective potential, it must correspond to $%
\chi =\pm \sqrt{n}a$, due to the orbit equation. By substituting one of
those values of $\chi $ in the potential $U_{eff}(\chi )$ and imposing that
they are minima of this last, a priori, positive definite effective
potential ($U_{eff}($ $\chi =\pm \sqrt{n}a)=0$), the critical value of $%
c_{0} $
\begin{equation}
c_{0}^{-1}=a^{n-2}\left[ \frac{n}{2}-1\right] n^{n/2}.  \label{eq24}
\end{equation}%
is obtained. Except for the case $n=1$ that presents two minima, one finds
positive definite effective potentials with three minima only for $n$ even.
Such kinds of polynomial potentials were already discussed extensively in
the literature \cite{lohe}. These polynomial potentials have typical \ kink
solutions that connect the minimum $\chi =0$ to one of the other two minima $%
\chi =\pm \sqrt{n}a$ and vice-versa. For $n=4$, we have for instance
\begin{equation}
\chi =\mp \sqrt{2}a\frac{\cosh (\mu ax)\pm \sinh (\mu ax)}{\sqrt{\cosh (2\mu
ax)}},  \label{eq25}
\end{equation}%
which have a kink profile very similar to that of the first-order
differential equation solutions for the effective models of the kind $U(\phi
)=(\phi ^{2}-a|\phi |)^{2}$. These solutions that can be seen as
half-torsion in a spin chain, also have similar profile to those exhibited
by self-consistent solutions for inhomogeneous chiral condensates in the
Nambu-Jona-Lasinio model in 1+1 space-time dimensions \cite{shei}. Since
these solutions are not continuously deformable into two-kinks solutions,
they will not be applied to the brane-world scenario considered below where
we are going to discuss the consequences of thicker branes in the warping of
the space.

\section{Application to a brane world scenario}

We now consider the scalar field coupled to gravity in (4,1) space-time
dimensions described by the action%
\begin{equation}
S=\int d^{4}xdy\sqrt{|g|}\left( -\frac{1}{4}R+\frac{1}{2}\partial _{a}\phi
\partial ^{a}\phi -V(\phi )\right) ,  \label{eq26}
\end{equation}%
where $g\equiv \mathrm{Det}(g_{ab})$ and the metric is
\begin{equation}
ds^{2}=g_{ab}dx^{a}dx^{b}=e^{2A(r)}\eta _{\mu \nu }dx^{\mu }dx^{\nu }-dr^{2},%
\hspace{0.5in}a,b=0,...,4,  \label{eq27}
\end{equation}%
where $r=x^{4}$ is the extra dimension, $\eta _{\mu \nu }$ is the Minkowski
metric and $e^{2A(r)}$ is the so-called warp factor, which is supposed to
depend only on the extra dimension. The Greek indices run from $0$ to $3$.

The static equations of motion following from the action (\ref{eq26}) and
for the case that the scalar field depends only on the extra dimension are
written as
\begin{equation}
\frac{d^{2}\phi }{dr^{2}}+4\frac{dA}{dr}\frac{d\phi }{dr}=\frac{dV(\phi )}{%
d\phi }  \label{eq28}
\end{equation}%
\begin{equation}
\frac{d^{2}A}{dr^{2}}=-\frac{2}{3}\left( \frac{d\phi }{dr}\right) ^{2}
\label{eq29}
\end{equation}%
and
\begin{equation}
\left( \frac{dA}{dr}\right) ^{2}=\frac{1}{6}\left( \frac{d\phi }{dr}\right)
^{2}-\frac{1}{3}V(\phi ).  \label{eq30}
\end{equation}

We consider that the potential $V(\phi )$ can be written as \cite{dewolfe}
\begin{equation}
V(\phi )=\frac{1}{2}\left( \frac{dW(\phi )}{d\phi }\right) ^{2}-\frac{4}{3}%
\left( W(\phi )\right) ^{2},  \label{eq31}
\end{equation}%
\noindent In this case the BPS\ solutions \cite{BPS} of the following
first-order differential equations
\begin{equation}
\frac{d\phi }{dr}=\pm \frac{dW(\phi )}{d\phi }\hspace{0.25in}\mathrm{and%
\hspace{0.25in}}\frac{dA}{dr}=\mp \frac{2}{3}W(\phi )  \label{eq32}
\end{equation}%
are also solutions of the second-order differential equations (\ref{eq28})
and (\ref{eq29}) and the equation (\ref{eq30}) is identically satisfied. By
taking $W_{\phi }(\phi )$ given by equation (\ref{eq17}) together with the
corresponding kink solution in (\ref{eq18}), the superpotential is given by
\begin{equation}
W(\phi )=2\mu \left[ \phi \left( \frac{\phi ^{2}}{3}+f^{2}+\frac{b}{2}\sqrt{%
\phi ^{2}+f^{2}}\right) +\frac{bf^{2}}{2}\sinh ^{-1}\left( \frac{\phi }{f}%
\right) \right] ,  \label{eq34}
\end{equation}%
and the warp factor is found by integrating the second of the equations (\ref%
{eq32}) with the classical solution (\ref{eq18}) substituted in $W(\phi )$.
We show in Figure 3 two profiles of the warp factor ($a=1$), corresponding
to two different values of the parameter $b$; one close to and the other one
far from the critical value $b=-1$. One can note that as far as $b$
decreases (its modulus increases) the warp factor becomes more narrow. For
values of $b$ close to the critical value one can observe that the warp
factor becomes wide and one can see a flat region, which signalizes a
Minkowskian metric inside the domain wall. Thick is the brane, wider is the
warp factor. We have also analyzed the case in which $W(\phi )$ given by
equation (\ref{eq20}) and have verified that the behavior of the warp factor
is not substantially different from that presented here.

Now we consider the stability of the system by analyzing the equations of
motion of linear small fluctuations around the classical solutions. This
issue is also important to realize the localization of the gravity inside
the domain wall \cite{gremm}-\cite{dewolfe}. This is done by means of a
perturbation of the metric, $ds^{2}=e^{2A(r)}(\eta _{\mu \nu }+\varepsilon
h_{\mu \nu })dx^{\mu }dx^{\nu }-dr^{2}$and a small perturbation around the
classical solution $\phi \rightarrow \bar{\phi}(r)+\varepsilon \widetilde{%
\phi }(r,x_{\mu })$, where $\varepsilon $ is a small number. By performing
those perturbations in the Lagrangian density and by expanding it up to $%
\mathcal{O(}\varepsilon )$, we obtain the following equations of motion%
\[
e^{-2A}\square \widetilde{\phi }-4\frac{dA}{dr}\frac{d\widetilde{\phi }}{dr}-%
\frac{d^{2}V}{d\phi ^{2}}\widetilde{\phi }=\frac{1}{2}\frac{d\phi }{dr}\eta
^{\mu \nu }\frac{dh_{\mu \nu }}{dr},
\]%
for the fluctuation of the scalar field and
\begin{eqnarray}
&&\left. -\frac{1}{2}\square h_{\mu \nu }+e^{2A}(\frac{1}{2}\frac{d}{dr}+2%
\frac{dA}{dr})\frac{dh_{\mu \nu }}{dr}-\frac{1}{2}\eta ^{\alpha \beta
}(\partial _{\mu }\partial _{\nu }h_{\alpha \beta }-\partial _{\mu }\partial
_{\nu }h_{\beta \nu }-\partial _{\nu }\partial _{\alpha }h_{\beta \mu
})+\right.  \nonumber \\
&&\left. +\frac{1}{2}\eta _{\mu \nu }e^{2A}\frac{dA}{dr}\partial _{r}(\eta
^{\alpha \beta }h_{\alpha \beta })+\frac{4}{3}e^{2A}\eta _{\mu \nu }\frac{dV%
}{d\phi }\widetilde{\phi }=0\right. ,  \label{eq36}
\end{eqnarray}%
for the fluctuations of the metric.

In general it is quite difficult to take into account linear fluctuations of
all components of the metric together with the fluctuations of the brane in
order to have a broad view of the linear stability of the whole system. This
is due to fact that the above set of coupled differential equations
involving the fluctuations is very intricate to be solved. Nevertheless, it
is possible to show that the transverse and traceless part of the
fluctuations of the metric ( $\overline{h}_{\mu \nu }$) decouple from
fluctuations of the brane \cite{dewolfe}. By constructing $\overline{h}_{\mu
\nu }=P_{\mu \nu \alpha \beta }h^{\alpha \beta }$ from the projector
operator $P_{\mu \nu \alpha \beta }\equiv \frac{1}{2}(\pi _{\mu \alpha }\pi
_{\nu \beta }+\pi _{\mu \nu }\pi _{\nu \alpha })-\frac{1}{3}\pi _{\mu \nu
}\pi _{\alpha \beta }$ with $\pi _{\mu \nu }\equiv \eta _{\mu \nu }-\partial
_{\mu }\partial _{\nu }/\square $, we have that the equation (\ref{eq36})
for the transverse and traceless part of the fluctuations of the metric
simplifies to
\begin{equation}
\frac{d^{2}\overline{h}_{\mu \nu }}{dr^{2}}+4\frac{dA}{dr}\frac{d\overline{h}%
_{\mu \nu }}{dr}-e^{-2A}\square \overline{h}_{\mu \nu }=0.  \label{eq37}
\end{equation}%
By using separation of variables and expressing $\overline{h}_{\mu \nu }$
conveniently as

\begin{equation}
\overline{h}_{\mu \nu }=e^{ik_{\mu }x^{\mu }}e^{-(3/2)A(r)}\xi _{\mu \nu
}(r),  \label{eq38}
\end{equation}%
we find, by making use of the transformation of variables $z=\int
e^{-A(r)}dr $, that $\xi _{\mu \nu }(z)$ satisfies the following Schr\"{o}%
dinger-like stability equation
\begin{equation}
-\frac{d^{2}\xi _{\mu \nu }}{dz^{2}}+V_{eff}(z)\xi _{\mu \nu }=k^{2}\xi
_{\mu \nu },  \label{eq39}
\end{equation}%
where
\begin{equation}
V_{eff}(z)=\frac{9}{4}\left( \frac{dA}{dz}\right) ^{2}+\frac{3}{2}\frac{%
d^{2}A}{dz^{2}}  \label{eq40}
\end{equation}%
is the effective potential. We show in the Figure 4 this effective potential
against the variable $r$ for two different values of $b$, one close and the
other one far from the critical value. One can note that the shape of the
effective potential for values of $b$ far from the critical value is similar
to the shape of others presented in the literature, for example in \cite%
{gremm}. Therefore, the only bound-state solution is the one associated to
the zero-mode which can be seen as the localization of the gravity inside
the domain wall. Higher energy modes are non-localized states that can
escape from the domain wall and propagate along the extra dimension. In any
case, no matter how big the value of $|b|$ is, there is no room to have
localized tachyon modes ($k^{2}<0$). The differential operator in equation (%
\ref{eq39}) can be factorized as the product of two operators which are
adjoint of each other:%
\begin{equation}
a^{\dag }a\xi _{\mu \nu }=\left( \frac{d}{dz}+\frac{3}{2}\frac{dA}{dz}%
\right) \left( -\frac{d}{dz}+\frac{3}{2}\frac{dA}{dz}\right) \xi _{\mu \nu
}=k^{2}\xi _{\mu \nu },  \label{eq41}
\end{equation}%
then, for the n-th normalized eigenmode $|n\rangle $ we have $%
k_{n}^{2}=\langle n|a^{\dag }a|n\rangle =|\left\vert a|n\rangle \right\vert
^{2}\geq 0$. Particularly, the non-normalized zero-energy eigenmode is given
by $\xi _{\mu \nu }^{(0)}(z)=e^{\left( 3/2\right) A(z)}$ and the
corresponding transverse and traceless part of the fluctuation of the metric
presents no dependence with the extra dimension.

Another aspect that the non-polynomial models share with other models is the
behavior of the matter energy density. It is given by%
\begin{equation}
\varepsilon (r)=e^{4A(r)}\left[ \frac{1}{2}\left( \frac{d~\bar{\phi}}{dr}%
\right) ^{2}+V(\bar{\phi})\right]  \label{eq42}
\end{equation}%
where $V(\bar{\phi})\ $is the potential in (\ref{eq31}) evaluated at the
classical solution. The behavior of \ the matter energy density is shown in
Figure 5 for two different values of $b$. The features shared with most of
the models is the peak of the energy density around the thick brane, which
can be observed for values of $b$ far from the critical value, and the
presence of regions outside the domain wall where the matter energy density
is negative. For values of $b$ close to the critical one, two relative small
peaks of the energy density show up around each wall of the double-wall
structure, this is also a common feature of the models with thicker branes
\cite{nosso, bazeia2}. Although the energy matter is negative, the energy
functional \cite{townsend}
\begin{equation}
F=\frac{1}{2}\int_{-\infty }^{+\infty }dr~e^{4A(r)}\left[ \left( \frac{d\phi
}{dr}\right) ^{2}+2V(\phi )-6\left( \frac{dA}{dr}\right) ^{2}\right] ,
\label{eq43}
\end{equation}%
is positive definite. This energy functional generates the Euler-Lagrange
equations (\ref{eq28})-(\ref{eq30}) and is minimized by the solutions of the
first-order differential equations (\ref{eq32}), since it can be rewritten as%
\begin{eqnarray}
F &=&\frac{1}{2}\int_{-\infty }^{+\infty }dr\left\{ ~e^{4A(r)}\left[ \left(
\frac{d\phi }{dr}\mp W_{\phi }\right) ^{2}-6\left( \frac{dA}{dr}\pm \frac{2}{%
3}W\right) ^{2}\right] \pm \frac{d}{dr}\left( e^{4A(r)}W\right) \right\} =
\nonumber \\
&=&\left. \pm e^{4A(r)}W(\bar{\phi})\right\vert _{-\infty }^{+\infty }.
\label{eq44}
\end{eqnarray}%
Note that the $W|_{\phi (\infty )}-W|_{\phi (-\infty )}$ is bigger (less)
than zero if we choose the upper (lower) sign. This energy plays the role of
the topological $E_{BPS}$ energy in the present scenario of scalar fields
interacting with gravitation in a (4,1) dimensional warped space-time.

\section{Final Remarks}

We have constructed effective models with only one scalar field which
supports deformed kink solutions which has been called two-kink solutions.
This has been done by means of a general orbit equation relating two scalar
fields and by eliminating one of them in terms of the other one. When such a
procedure is done in the Lagrangian density, the effective Lagrangian
density is not given in the usual kinetic$-$potential\ form (see equations (%
\ref{eq5})-(\ref{eq5a})). In order to have an effective canonical Lagrangian
density we adopt the elimination of one of the field in the equations of
motion, instead. The model we have constructed here together with their
two-kinks solutions are very similar to the one propose in \cite{dutra09},
but, in principle, other (polynomial) models are obtained from that one we
have started with. As a matter of fact, a myriad of models with two scalar
interacting fields have been studied in reference \cite{bazeia5} and from
those we can, in principle, construct other effective models with only one
scalar field. One of the models which is under analysis is the fourth model
proposed in \cite{bazeia5}, whose potential is given by $V(\phi ,\chi )=\bar{%
\lambda}\phi +\bar{\mu}\chi -\lambda /4(\phi ^{4}+\chi ^{4}+6\phi ^{2}\chi
^{2})$.

One of the effective models constructed here is applied to a brane-world
scenario where its influence in the warp factor leads to a flat geometry
inside the thicker domain walls. That model can also be used to analyze
universal aspects of thick branes splitting in a warped geometry \cite%
{campos}. That phenomenon can be interpreted as a phase transition due to
the variation of the parameters of the potential, which in our case are the
parameters $b$ for the superpotential (\ref{eq17}), and $c$ for the
superpotential (\ref{eq20}). In each case there will be a transition in the
form of the potential, from one with two vacua to another one with three
vacua when the parameter reaches a critical value. One can think of $b$, for
instance, as dependent on the temperature. Far from the critical temperature
we have a single relatively thick brane and as the critical temperature is
approached, the parameter $b$ approaches $-a$ from the left. For values of $%
b $ close to but less than $-a$ the thick brane splitting starts to take
place (note the behavior of the solid curve in Figure 2) and the local
minimum in the effective potential $U_{eff}$ presents the tendency to become
a global minimum. The splitting also influences the localization of the zero
modes fluctuation of the metric as can be seen by the behavior of the
effective potential in Figure 5. In that figure one can see that for values
of $b$ far from the critical value the effective potential has a volcano
shape which leads to a narrow (more localized) zero-mode fluctuation than in
the situation in which $b$ is close to the critical value. The splitting of
the brane is also manifested in the splitting of the matter energy density
shown in Figure 6. When $b=-a$, the effective potential is $U_{eff}(\phi
)=2\mu ^{2}(\phi ^{2}-a|\phi |)^{2}\,$and the brane is split in two. We can
suppose that the branes are at a distance $2L$ from each other and localized
around the core of each one of the solutions $\phi _{-}(r)=-a/2\{1-\tanh
[2\mu a(r+L)]\}$ and $\phi _{+}(r)=a/2\{1+\tanh [2\mu a(r-L)]$. We have not
analyzed what happens to the warp factor and the fluctuation of the metric
under the influence of both scalar fields simultaneously but this might be
done by following the same numerical approach adopted in the reference \cite%
{vacha}, where it was obtained the spectrum of fermions in the background of
a kink and a anti-kink which are far apart from each other. We think that
the effective potential for the fluctuations of the metric will have the
shape of two volcanos whose craters are distant $2L$ from each other. That
effective potential can support a zero eigenmode and the eigenfunctions may
have peaks around each brane or in the region between the branes.

The effective models presented here can also be considered in
space-time with $D>4$ dimensions and with the solutions depending
only on the radial coordinate. Since nonlinear models with only
scalar fields are not stable in space-time dimensions bigger than
two, as has been demonstrated by Derricks theorem \cite{derrick},
one should resort to a convenient bypass by introducing an explicit
dependence of the interacting potential on the coordinates, as has
been provided by \cite{bazeia3}. The resulting radial solutions are
very similar to those shown in the reference \cite{dutra09}. Those
solutions and their consequences in warped space-time with two and
three extra dimensions is under study and the results will be
reported elsewhere.

\bigskip \bigskip

\textbf{Acknowledgments}

We are grateful to A. de Souza Dutra for helpful discussions on matters
concerning brane-worlds and many suggestions on technical details. We also
thank to F. A. Brito for calling us attention to the reference \cite%
{townsend}. This work has been partially financed by CAPES and CNPq.

\newpage

\begin{figure}[tbp]
\begin{center}
\epsfig{file=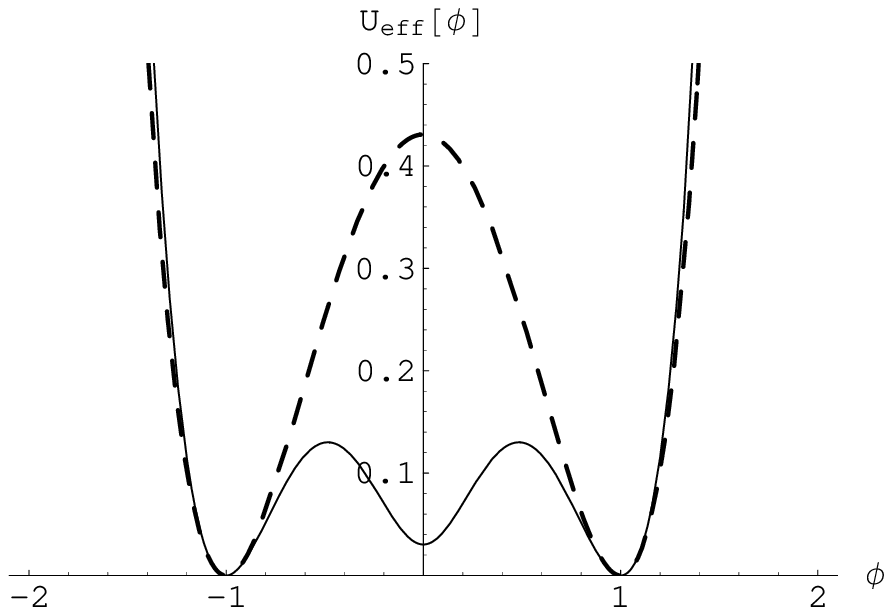}
\end{center}
\par
\caption{The effective potential in the case $\protect\lambda=\protect\mu$,
for $a=1$ and $b=-1.001$ (solid line) and $b=-1.3$ (dashed line).}
\end{figure}

\begin{figure}[tbp]
\begin{center}
\epsfig{file=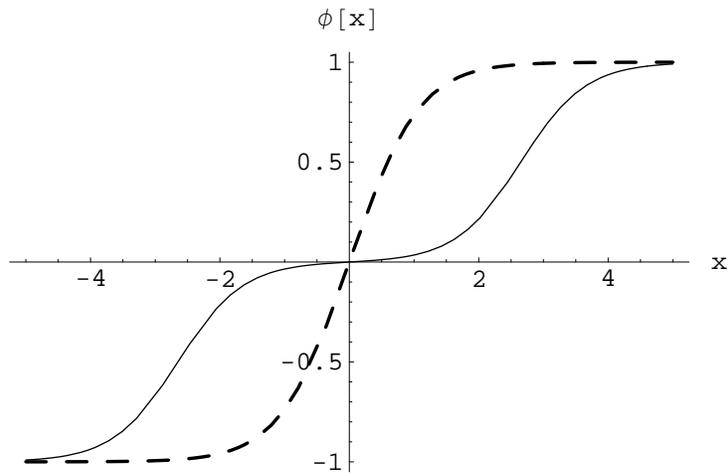}
\end{center}
\par
\caption{Typical kink-profiles, for $a=1$ and $b=-1.001$ (solid line),
corresponding to two-kinks solution and $b=-1.3$ (dashed line),
corresponding to a single kink solution.}
\end{figure}

\begin{figure}[tbp]
\begin{center}
\epsfig{file=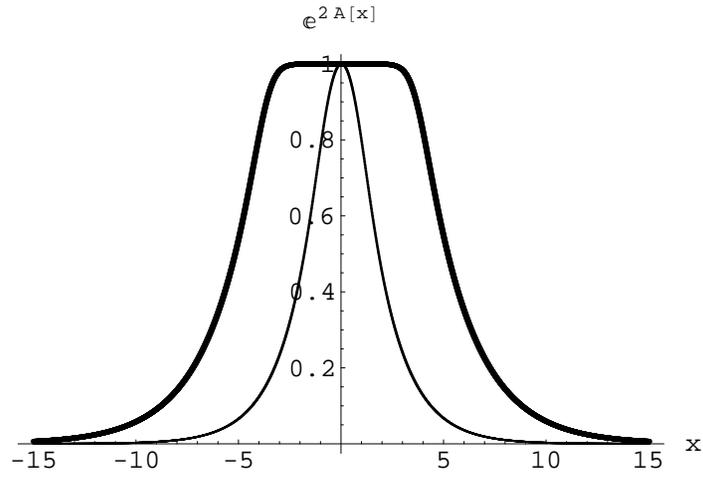}
\end{center}
\par
\caption{Warp factor for two different values of $b$. One closed to
the critical value (thick line) with a \textit{meseta} shape and the
other one far from the critical (thin line).}
\end{figure}
\begin{figure}[tbp]
\begin{center}
\epsfig{file=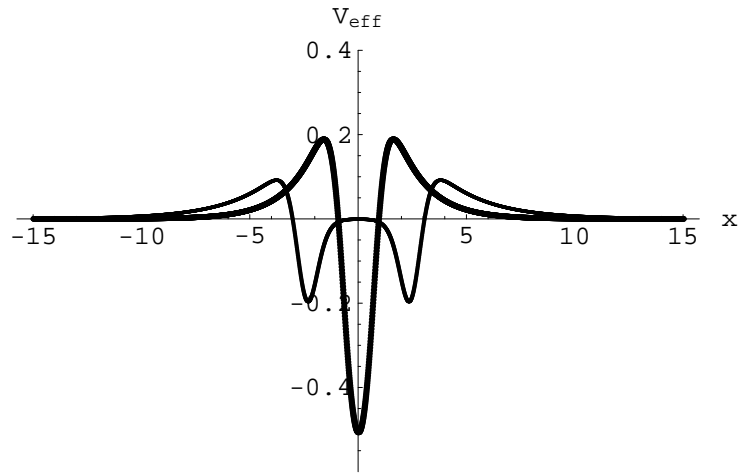}
\end{center}
\par
\caption{The effective potential in the effective Schr\"{o}dinger equation
for the fluctuation of the metric for two different values $b$, $b=-1.0001$
(thick line) and $b=-1.2$ (thin line).}
\end{figure}
\begin{figure}[tbp]
\begin{center}
\epsfig{file=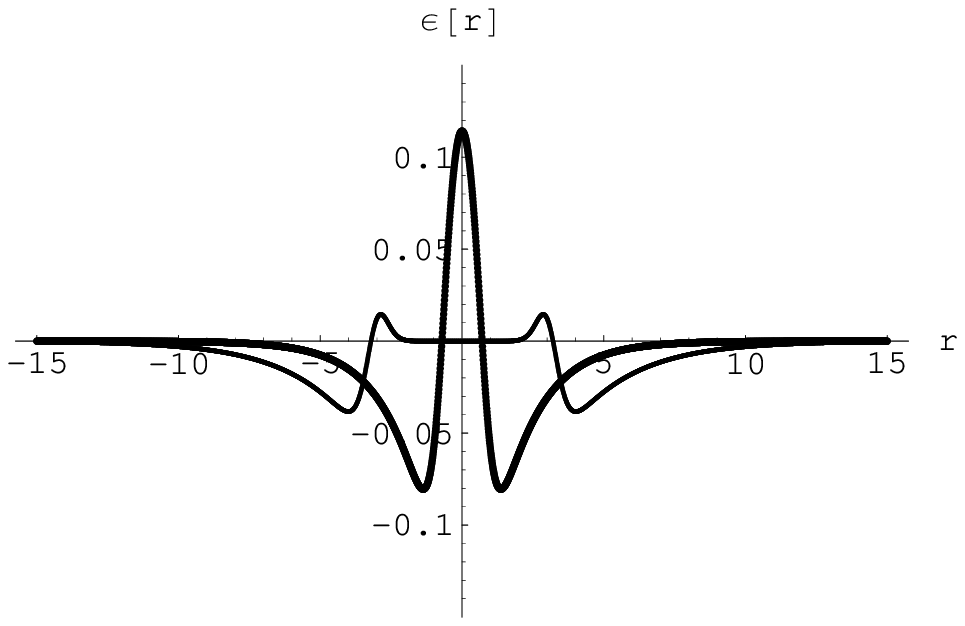}
\end{center}
\par
\caption{The matter energy density in the case $\protect\lambda=\protect\mu$%
, for $a=1$ and $b=-1.0001$ (thin line) and $b=-1.3$ (thick line).}
\end{figure}


\bigskip

\begin{thebibliography}{99}
\bibitem{randallsundrum} L. Randall, R. Sundrum, Phys. Rev. Lett 83 (1999)
3370; Phys. Rev. Lett. 83 (1999) 4690.

\bibitem{gremm} M. Gremm, Phys. Lett. B 478 (2000) 434.

\bibitem{kyrgiz} For a recent review on thick branes see: V.
Dzhunushaliev,V. Folomeev, M. Minamitsuji, \textit{Thick brane solutions, }%
arXiv:gr-qc/0904.1775v2.

\bibitem{campos} A.Campos, Phys.Rev. Lett. 88\textbf{\ \ }(2002) 141602
(2002).

\bibitem{bazeia1} D. Bazeia, A. R. Gomes, JHEP 05 (2004) 012.

\bibitem{nosso} A. de Souza Dutra, A. C. Amaro de Faria Jr., M. B. Hott,
Phys. Rev. D 78 (2008) 043526.

\bibitem{dutra09} A. de Souza Dutra, Physica D 238 (2009) 798.

\bibitem{bazeia2} D. Bazeia, J. Furtado, A. R. Gomes, JCAP 02 (2004) 002.

\bibitem{bazeia3} D. Bazeia, J. Menezes, R. Menezes, Phys. Rev. Lett. 91
(2003) 241601.

\bibitem{dutra05} A. de Souza Dutra, Phys. Lett. B 626 (2005) 249.

\bibitem{BPS} M. K. Prasad, C. M. Sommerfeld, Phys. Rev. Lett. 35 (1975)
760. E. B. Bolgomol 'nyi, Sov. J. Nucl. Phys. 24 (1976) 449.

\bibitem{dutra04} A. de Souza Dutra, A. C. Amarao de Faria Jr., Czech. J.
Phys. 54 (2004) 1229.

\bibitem{bazeia4} D. Bazeia, L. Losano, J. M. C. Malbouisson, Phys. Rev. D
66 (2002) 101701(R).

\bibitem{bazeia0} D. Bazeia, M.J. dos Santos, R.F. Ribeiro, Phys. Lett A 208
(1995) 84. D. Bazeia, R. F. Ribeiro, M. M. Santos, Phys. Rev. E 54 (1996)
2943.

\bibitem{lohe} M. A. Lohe, Phys. Rev. D 20 (1979) 3120. M. A. Lohe, D. M.
O'Brien, Phys. Rev. D (1981) 1771.

\bibitem{shei} S. S. Shei, Phys. Rev. D 14 (1976) 535.

\bibitem{dewolfe} O. DeWolfe, D. Z. Freedman, S. S. Gubser, A. Karch, Phys.
Rev. D 62 (2000) 046008. O. DeWolfe, J. Erlich, C. Gorjean, T.J. Hollowood,
Nucl. Phys. B 584 (2000) 359.

\bibitem{townsend} K. Skenderis, P. K. Townsend, Phys. Lett. B 468 (1999) 46.

\bibitem{bazeia5} D. Bazeia, F. A. Brito, Phys. Rev. D 61 (2000) 105019.

\bibitem{vacha} Yi-Zen Chu, T. Vachaspati, Phys. Rev. D 77\ (2008) 025006.

\bibitem{derrick} G. H. Derrick, J. Math. Phys. 5 (1964) 1252.
\end{thebibliography}
\end{document}